\documentstyle[prl,aps,multicol]{revtex}

\def\beq{\begin{equation}}
\def\eeq{\end{equation}}

\renewcommand{\narrowtext}{\begin{multicols}{2} \global\columnwidth20.5pc}
\renewcommand{\widetext}{\end{multicols} \global\columnwidth42.5pc}
\begin{document}


\vspace{0.5cm}

\title{  Bi-maximal Neutrino Mixings And Proton Decay \\ 
In $SO(10)$ With Anomalous Flavor ${\cal U}(1)$}

\author{Qaisar Shafi$^a$
and Zurab Tavartkiladze$^b$
}

\address{{\it $^a$ Bartol Research Institute, University of Delaware,
Newark, DE 19716, USA  \\
$^b$Institute of Physics, Georgian Academy of Sciences,
380077 Tbilisi, Georgia }}

\maketitle

\begin{abstract}
By supplementing supersymmetric $SO(10)$ with an anomalous ${\cal U}(1)$
flavor symmetry and additional `matter' superfields carrying suitable
${\cal U}(1)$ charges, we
explain the charged fermion
mass hierarchies, the magnitudes of the CKM matrix elements, as well as the
solar and atmospheric neutrino data. We stress bi-maximal vacuum neutrino
mixings, and indicate how the small or large mixing angle MSW solution can
be incorporated. The ${\cal U}(1)$
symmetry also implies that
$\tau_{p\to K\nu }[SO(10)]\sim (10-100)\cdot
\tau_{p\to K\nu}[{\rm minimal}~SU(5)]$.
\vspace*{-6.5cm}
\begin{flushright}
BA-99-63
\end{flushright}
\vspace*{6.5cm}
\end{abstract}



 \narrowtext
            
The possible existance of neutrino oscillations indicated 
by the solar and
atmospheric neutrino data \cite{atmSK} has inspired many recent attempts 
to incorporate the
phenomena within a variety of theoretical frameworks
\cite{bimax}-\cite{maxmix}. 
Of course, realistic scenarios should also shed light on the charged
fermion
masses and their mixings, and we will attempt to do that in this letter. 
Our main focus
is to realize bi-maximal
neutrino mixings \cite{bimax}-\cite{bimaxso10} 
within the $SO(10)$ framework, generalizing our
earlier work based on $SU(5)$ \cite{bimaxsu5}. The mechanism we propose 
differs 
from those suggested earlier for $SO(10)$ \cite{deg2}, \cite{bimaxso10} 
and turns out to have important
consequences for proton decay as well. An important role in our considerations
is played by an anomalous flavor ${\cal U}(1)$ symmetry \cite{anu1},
which helps realize the mechanism of maximal neutrino mixings 
\cite{maxmix}.
We also indicate how, if future data requires it, the small or large
mixing angle MSW solutions \cite{solSK} can be incorporated.

For the symmetry breaking $SO(10)$ to
$SU(3)_c\times SU(2)_L\times U(1)_Y\equiv G_{321}$, the `minimal' 
higgs system consists of $45+\overline{16}+16$. 
However, many $SO(10)$ scenarios \cite{DT} 
invoke several
$45$ or $54$-plets, or both. Without studying this issue in detail, we will
let $\Sigma $ denote the $45$ and/or $54$ multiplets, and $\bar C+C$ the
$\overline{16}+16$ higgs respectively.
Their VEVs are responsible for breaking $SO(10)$  down to
$G_{321}$. We will assume that the doublet fragments which reside in 
the higgs $10$-plet($\equiv H$) are light, while 
the color triplets in $H$ acquire masses of the
order of GUT scale($\equiv M_G\sim 2\cdot 10^{16}$~GeV) 
(such a desirable doublet-triplet
splitting may be achieved through one of the mechanisms suggested earlier
\cite{DT}). For building a realistic `matter' sector we extend
the higgs sector with an $SO(10)$ singlet $S$ and
$\overline{16}+16$-plets$\equiv \bar C'+C'$ 
(these states allow us to select the transformation 
properties of the various superfields under the additional symmetries,
see below, in such a 
way as to forbid all nondesirable operators in our scenario). The GUT
scale VEVs are 
\beq
\langle \bar C\rangle=\langle C\rangle \sim \langle \Sigma \rangle
\sim \langle S\rangle
\sim M_G\equiv M_P\epsilon_G,~~~\epsilon_G\sim 10^{-2}~,   
\label{gutvevs}
\eeq
while $\langle \bar C'\rangle=\langle C'\rangle=0$,
where $M_P\simeq 2.4\cdot 10^{18}$~GeV denotes the reduced Planck scale.

We next turn to the symmetries of the theory.
Together with `matter' parity, we introduce 
a ${\cal Z}_2$ ${\cal R}$-symmetry
under which the various superfields and superpotential
$W$ have the following transformation properties:
$$
\left(\Sigma, S, H, \bar C, C, C'\right)\to
-\left(\Sigma, S, H, \bar C, C, C'\right)~,
$$
\beq
\bar C'\to \bar C'~,~~~~W\to -W~.
\label{scalz2}
\eeq
The ${\cal Z}_2$ symmetry  
is crucial for a `natural' generation
of appropriate mass scales. 

For understanding the hierarchies between
charged fermion masses and the magnitudes of the CKM matrix elements, 
we introduce an anomalous
${\cal U}(1)$ symmetry which often emerges from strings. The
associated Fayet-Iliopoulos term equals \cite{fi}
\beq
\xi \int d^4\theta V_A~,~~~~
\xi =\frac{g_A^2M_P^2}{192\pi^2}{\rm Tr}Q~.
\label{fi}
\eeq
The $D_A$-term is
\begin{equation}
\frac{g_A^2}{8}D_A^2=\frac{g_A^2}{8}
\left(\Sigma Q_a|\varphi_a |^2+\xi \right)^2~,
\label{da}
\end{equation}
where $Q_a$ is the `anomalous' charge of $\varphi_a $ superfield.

We will introduce an $SO(10)$ singlet superfield $X$ with ${\cal U}(1)$
charge $Q_{X}=1$, whose VEV breaks ${\cal U}(1)$.
Assuming ${\rm Tr}Q<0$ ($\xi <0$), the cancellation of (\ref{da})
fixes the VEV of the scalar component of $X$:
\beq
\langle X\rangle =\sqrt{-\xi }~.
\label{vevx}
\eeq
We will assume that
\beq
\frac{\langle X\rangle }{M_P}\equiv \epsilon \simeq 0.22~,
\label{epsx}
\eeq
where the parameter $\epsilon $ plays an essential role in our analysis.
By exploiting the anomalous ${\cal U}(1)$  
as a flavor symmetry \cite{anu1}, we will gain a natural understanding
of the hierarchies of the charged fermion masses and mixings, whose 
magnitudes will be
expressed in powers of $\epsilon $ (under
${\cal Z}_2$ symmetry, $X\to X$).

The ${\cal U}(1)$ charges of the `higgs' superfields are assigned to be
$$
Q_{X}=1~,~~~Q_{\Sigma }=Q_S=0~,~~~Q_H=-5/2
$$
\beq
Q_{C}=-Q_{\bar C}=-15/4~,~~Q_{\bar C'}=-5/4~,~~
Q_{C'}=-15/4~.
\label{scalu1}
\eeq

{}From the superpotential couplings 
\beq
\frac{\Sigma +S}{M_P}H\bar C\bar C'+
\left(\frac{X}{M_P} \right)^5M_P\bar C'C'~,
\label{scal1}
\eeq
one can readily verify that the `light' $h_u$ fully resides
in $H$, while $h_d$ resides in $H$ and $C'$, with `weights'
of order unity and $\frac{\epsilon_G^2}{\epsilon^5}\sim 0.2$ respectively:
\beq
H\supset (h_u, h_d)~,~~~C'\stackrel{\supset }{_\sim }
\frac{\epsilon_G^2}{\epsilon^5}h_d~.
\label{weights} 
\eeq

The Yukawa sector, constructed with the minimal set of `matter'
$16_i$'s ($i=1, 2, 3$), involve the couplings $16_i16_jH$, 
which yield the undesirable
asymptotic relations $\hat{m}_U^0=\hat{m}_D^0=\hat{m}_E^0$, and a trivial CKM
matrix $\hat{V}_{CKM}={\bf 1}$. For obtaining a realistic pattern of fermion
masses and mixings, we extend the `matter' sector and introduce 
three supermultiplets $10_i$ (which
are crucial for removing the $\hat{m}_U^0=\hat{m}_{D, E}^0$ degeneracy
\cite{deg1}, \cite{deg2}), and two pairs of vector-like states 
$[\bar F(\overline{16})+F(16)]_{1, 2}$ (which help avoid 
the relation $\hat{m}_D^0=\hat{m}_E^0$). 
The transformation properties of the various
`matter' superfields are:
$$
{\cal U}(1)~:~
~~Q_{16_1}=-7/4~,~~Q_{16_2}=-3/4~,~~
Q_{16_3}=5/4~,~~~~~~
$$
$$
Q_{10_1}=3/2~,~~Q_{10_2}=Q_{10_3}=5/2~,~~
$$
\beq
Q_{F_1}=-Q_{\bar F_1}=Q_{16_1}~,~~Q_{F_2}=-Q_{\bar F_2}=Q_{16_2}~,
\label{feru1}
\eeq                                         
\beq
{\cal Z}_2: \left(16_i, 10_i, F_{1, 2}, \bar F_{1, 2} \right)\to
-\left(16_i, 10_i, F_{1, 2}, \bar F_{1, 2} \right).
\label{ferz2}
\eeq
{}From the couplings
\begin{equation}
\begin{array}{ccc}
&  {\begin{array}{ccc}
\hspace{-5mm}~~10_1 & \,\,10_2  & \,\,10_3

\end{array}}\\ \vspace{2mm}
\begin{array}{c}
16_1 \\ 16_2 \\16_3
 \end{array}\!\!\!\!\! &{\left(\begin{array}{ccc}
\,\,\epsilon^4~~ &\,\,\epsilon^3~~ &
\,\,\epsilon^3
\\
\,\,\epsilon^3~~   &\,\,\epsilon^2~~  &
\,\,\epsilon^2
 \\
\,\,\epsilon~~ &\,\,1~~ &\,\,1
\end{array}\right)C }~,
\end{array}  \!\!  ~~~~~
\label{massive}
\end{equation}
after substituting the VEV of $C$, the $\bar 5_i$ fragments (of $SU(5)$),
which reside in $16_i$-plets form heavy massive states with $5_{10_i}$.
Since the couplings $10_i10_j$ are forbidden, the 
`light' states $(d^c, l)_i$ fully reside in $10_i$. 
The remaining `light' quark-lepton
states reside in $16_i$. We therefore 
have                                    
\beq
10_i\supset (d^c, l)_i~,~~~~16_i\supset (q, u^c, e^c)_i~.
\label{weightfer}
\eeq 

The couplings
\begin{equation}
\begin{array}{ccc}
 & {\begin{array}{ccc}
\hspace{-5mm}~~16_1 & \,\,16_2  & \,\,16_3   

\end{array}}\\ \vspace{2mm}
\begin{array}{c}
16_1 \\ 16_2 \\16_3
 \end{array}\!\!\!\!\! &{\left(\begin{array}{ccc}
\,\,\epsilon^6~~ &\,\,\epsilon^5~~ &
\,\,\epsilon^3
\\
\,\,\epsilon^5~~   &\,\,\epsilon^4~~  &
\,\,\epsilon^2
 \\
\,\,\epsilon^3~~ &\,\,\epsilon^2~~ &\,\,1
\end{array}\right)H }~,
\end{array}  \!\!  ~~~~~
\label{up}
\end{equation}
which respect the transformation properties 
(\ref{scalz2}), (\ref{scalu1}), (\ref{feru1}),
(\ref{ferz2}), are responsible for generating masses for the
up type quarks. Taking into account (\ref{weights}),
(\ref{weightfer}) yields the desirable hierarchies
\beq
\lambda_t\sim 1~,~~~\lambda_u : \lambda_c :  \lambda_t \sim
\epsilon^6 : \epsilon^4 :1~.
\label{lambdaup}
\eeq

The down quark and charged lepton masses emerge from the couplings
\begin{equation}
\begin{array}{ccc}
 & {\begin{array}{ccc}
\hspace{-5mm}~~10_1 & \,\,10_2  & \,\,10_3

\end{array}}\\ \vspace{2mm}
\begin{array}{c}
16_1 \\ 16_2 \\16_3
 \end{array}\!\!\!\!\! &{\left(\begin{array}{ccc}
\,\,\epsilon^4~~ &\,\,\epsilon^3~~ &
\,\,\epsilon^3
\\
\,\,\epsilon^3~~   &\,\,\epsilon^2~~  &
\,\,\epsilon^2
 \\
\,\,\epsilon~~ &\,\,1~~ &\,\,1
\end{array}\right)C' }~.
\end{array}  \!\!  ~~~~~
\label{downlep}
\eeq
Equations (\ref{weights}), (\ref{weightfer}) yield the
desirable hierarchies
$$
\lambda_b\sim \frac{\epsilon_G^2}{\epsilon^5}~,~~
\lambda_d :\lambda_s :\lambda_b \sim
\epsilon^4:\epsilon^2 :1~,~~~
\tan \beta \sim\frac{\epsilon_G^2}{\epsilon^5}\frac{m_t}{m_b}~,
$$
\beq
\lambda_{\tau }\sim \frac{\epsilon_G^2}{\epsilon^5}~,~~
\lambda_e :\lambda_{\mu } :\lambda_{\tau } \sim
\epsilon^4:\epsilon^2 :1~.
\label{lambdas}
\eeq

However, the degeneracy $\hat{m}_D^0=\hat{m}_E^0$ still holds at this stage
since the $SU(5)$ symmetry is not broken in (\ref{downlep}). 
To remove this
drawback, we invoke the states $(\bar F+F)_{1, 2}$. 
The relevant couplings will be
\begin{equation}
\begin{array}{cc}
 & {\begin{array}{ccc}
16_1&\,\,16_2&\,\,16_3~~~~~~~~~~~~
\end{array}}\\ \vspace{2mm}
\begin{array}{c}
\bar F_1\\ \bar F_2

\end{array}\!\!\!\!\! &{\left(\begin{array}{ccc}
\,\, 1~~&
\,\,  0~~ &\,\, 0
\\
\,\, \epsilon ~~ &\,\,1~~&\,\, 0~
\end{array}\right)(S+\Sigma) }~,
\end{array}  \!\!~ 
\begin{array}{cc}
 & {\begin{array}{cc}
F_1&\,\,~F_2~~~~~~~~~~~~
\end{array}}\\ \vspace{2mm}
\begin{array}{c}
\bar F_1 \\ \bar F_2

\end{array}\!\!\!\!\! &{\left(\begin{array}{ccc}
\,\, 1~~
 &\,\,0
\\
\,\, \epsilon~~
&\,\,1 
\end{array}\right)(S+\Sigma).
}
\end{array}
\label{F16}
\end{equation}
{}From (\ref{F16}) we verify that the mixings of $(q, u^c, e^c)_{16_{1, 2}}$
with $(q, u^c, e^c)_{F_{1, 2}}$ are of the same order as the masses of
$(\bar F+F)_{1, 2}$ states. This means that the light states 
$(q, u^c, e^c)_{1, 2}$ remain with `weights' of order unity in $16_{1, 2}$
and $F_{1, 2}$ respectively. Note that the $\Sigma $ field(s) violate $SU(5)$
in (\ref{F16}), and therefore the unwanted asymptotic relations
$m_e^0=m_d^0$, $m_{\mu }^0=m_s^0$ are avoided, while the hierarchical
structure in (\ref{lambdas}) is unchanged. Since the states in
$16_3$ are not affected by (\ref{F16}), $b-\tau $
unification still holds at the GUT scale.

{}From (\ref{up}) and (\ref{downlep}), for the CKM matrix elements we find
\beq
V_{us} \sim \epsilon \ , \ V_{cb} \sim \epsilon^2 \ ,\ V_{ub} \sim
\epsilon^3~.
\label{ckm}
\eeq
Thus, thanks to the anomalous ${\cal U}(1)$ flavor symmetry and
our choice of extended `matter' multiplets, one can obtain a 
realistic pattern for the charged fermion masses
and CKM mixings within the framework of SUSY $SO(10)$.  

Next we turn to the neutrino sector and attempt to account for the solar
and atmospheric neutrino data. We concentrate on the
bi-maximal (vacuum) mixing scenario, but later point out how 
the small (or large) mixing
angle MSW oscillations can be realized in the present framework. 
Since the light fragments of $l_i$ reside in $10_i$ 
states, and $10_2$ and $10_3$ have the same ${\cal U}(1)$ charge 
(see (\ref{feru1})), we can expect naturally large $\nu_{\mu }-\nu_{\tau }$
mixing. This also can be seen from the texture in (\ref{downlep}). 
Introducing an $SO(10)$ singlet right handed neutrino ${\cal N}_3$
with suitable mass, the state
`$\nu_3$' will gain an appropriate mass relevant for the atmospheric
puzzle. At this stage the other two neutrino states are massless. 
{}From (\ref{downlep}) one can see that large $\nu_e-\nu_{\mu, \tau}$ mixing
will not be realized in a straightforward way (expected mixing 
is of order $\epsilon $). 

To obtain large 
$\nu_e-\nu_{\mu, \tau}$ mixing, we invoke the mechanism suggested in
\cite{maxmix}, which naturally yields `maximal' mixings between neutrino
flavors.
For this purpose we introduce two additional $SO(10)$ singlet states
${\cal N}_1$, ${\cal N}_2$. The states 
${\cal N}_1, {\cal N}_2, {\cal N}_3$ have the following transformation 
properties under ${\cal U}(1)\times {\cal Z}_2$:
$$
{\cal U}(1)~:~Q_{{\cal N}_1}=-Q_{{\cal N}_2}=-2~,~~~Q_{{\cal N}_3}=0~,
$$
\beq
{\cal Z}_2~:~~~~{\cal N}_i\to -{\cal N}_i~.
\label{Nu1z2}
\eeq
The relevant couplings are 
\beq
W_{{\cal N}_3}= \kappa S{\cal N}_3^2+
\epsilon (a \epsilon 10_1+b10_2+c10_3)H{\cal N}_3 ,
\label{N3}
\eeq
\begin{equation}
\begin{array}{cc}
 & {\begin{array}{cc}
{\cal N}_1~&\,\,{\cal N}_2~~~~~~
\end{array}}\\ \vspace{2mm}
\begin{array}{c}
10_1\\ 10_2 \\ 10_3

\end{array}\!\!\!\!\! &{\left(\begin{array}{ccc}
\,\, \epsilon^4~~ &
\,\,  1
\\
\,\, \epsilon^3~~ &\,\,0   
\\
\,\, \epsilon^3~~ &\,\,0
\end{array}\right)H }~,
\end{array}  \!\!~~~
\begin{array}{cc} 
 & {\begin{array}{cc}   
{\cal N}_1~&\,\,
{\cal N}_2~~~~~
\end{array}}\\ \vspace{2mm} 
\begin{array}{c}   
{\cal N}_1 \\ {\cal N}_2

\end{array}\!\!\!\!\! &{\left(\begin{array}{ccc}
\,\, \epsilon^4
 &\,\,~~~1
\\
\,\, 1
&\,\,~~~0
\end{array}\right)\kappa'S~,
}
\end{array}~~~
\label{Ns}
\end{equation}  
where $\kappa ,\kappa', a, b, c$ are dimensionless coefficients.
Note that there also exists the coupling
$\alpha S\epsilon^2{\cal N}_1{\cal N}_3$ which, if properly suppressed
(see (\ref{cond})), will not be relevant. 

Let us choose the basis in which the charged lepton matrix (\ref{downlep})
is diagonal. This choice is convenient because the matrix which
diagonalizes the neutrino mass matrix will then coincide 
with the lepton mixing
matrix. The hierarchical structure of the couplings in (\ref{N3})
will not be altered, while the `Dirac' and `Majorana' masses from (\ref{Ns})
will respectively have the forms
\begin{equation}
\begin{array}{cc}
m_D=\!\!\!\!\! &{\left(\begin{array}{ccc}
\,\, \epsilon^4~ &
\,\,  1
\\
\,\, \epsilon^3~ &\,\,\epsilon
\\
\,\, \epsilon^3~ &\,\,\epsilon
\end{array}\right)h_u }~,
\end{array}
\begin{array}{cc}

~~M_R=\!\!\!\!\! &{\left(\begin{array}{ccc}
\,\, \epsilon^4
 &\,\,1
\\
\,\, 1
&\,\,0
\end{array}\right)M_P\kappa'\epsilon_G~.
}
\end{array}
\label{dirmaj}
\end{equation}

Taking 
\beq
\kappa \sim 10^{-3}~,~~~~ \alpha < 2\cdot 10^{-2}
\label{cond}
\eeq
and the other coefficients of order unity, integration of the 
${\cal N}$ states leads to the following `light' neutrino mass matrix:
\beq
\hat{m}_{\nu }=\hat{A}m+\hat{B}m'~,
\label{matnu}
\eeq
where
\beq
m\equiv \frac{\epsilon^2h_u^2}{\kappa M_P\epsilon_G}~,~~~~
m'\equiv\frac{\epsilon^3h_u^2}{M_P\kappa'\epsilon_G}~,
\label{scales}
\eeq
$$
\begin{array}{ccc}
 & {\begin{array}{ccc}
~& \,\,~  & \,\,~~
\end{array}}\\ \vspace{2mm}
\hat{A}=
\begin{array}{c}
\\  \\
 \end{array}\!\!\!\!\!\!\!\!&{\left(\begin{array}{ccc}
\,\,a^2\epsilon^2  &\,\,~~ab\epsilon &
\,\,~~ac\epsilon
\\
\,\,ab\epsilon   &\,\,~~b^2  &
\,\,~~bc
 \\
\,\, ac\epsilon &\,\,~~bc  &\,\,~~c^2
\end{array}\right)m }~, 
\end{array}  \!\!  ~~
$$
\beq
\begin{array}{ccc} 
 & {\begin{array}{ccc}  
~~\,\,~  & \,\,~~~~
\end{array}}\\ \vspace{2mm}
\hat{B}=
\begin{array}{c}
\\  \\
 \end{array}\!\!\!\!\!\!\!\!&{\left(\begin{array}{ccc}
\,\,\epsilon &\,\,~1 &
\,\,~1
\\
\,\,1   &\,\,~\epsilon  &
\,\,~\epsilon
 \\
\,\, 1 &\,\,~ \epsilon  &\,\,~\epsilon
\end{array}\right)m' }
\end{array}  \!\!  ~.
\label{AB}
\end{equation}

The `light' eigenvalues are
$$  
m_{\nu_3}\simeq m(b^2+c^2+a^2\epsilon^2)
\sim 6\cdot 10^{-2}~{\rm eV}~,
$$  
\beq
m_{\nu_1 }\simeq m_{\nu_2 }\simeq m'\sim 1.3\cdot 10^{-5}~{\rm eV}~.
\label{masses}
\eeq
Ignoring CP violation the neutrino mass matrix (\ref{matnu})
can be diagonalized by the
orthogonal transformations $\nu_{\alpha }=U_{\nu}^{\alpha i}\nu_i$, where
$\alpha =e, \mu, \tau $ denotes flavor indices, and $i=1, 2, 3$
the mass eigenstates.
$U_{\nu }$ has the form
\beq
\begin{array}{ccc}
U_{\nu }=~~
\!\!\!\!\!\!\!\!&{\left(\begin{array}{ccc}
\,\,\frac{1}{\sqrt{2}} &\,\,~~\frac{1}{\sqrt{2}} &
\,\,~~s_1
\\
\,\,-\frac{1}{\sqrt{2}}c_{\theta }  &\,\,~~~~\frac{1}{\sqrt{2}}c_{\theta }
&\,\,~~s_{\theta }
\\
\,\, ~~\frac{1}{\sqrt{2}}s_{\theta } &\,\,~
-\frac{1}{\sqrt{2}}s_{\theta }  &\,\,~~c_{\theta }
\end{array}\right) }
\end{array}~,
\label{lepckm}
\end{equation}
with
\beq
\tan \theta =\frac{b}{c}~,~~~~s_1=\frac{a\epsilon}{\sqrt{b^2+c^2}}~,
\label{angles}
\eeq
and $s_{\theta }\equiv \sin \theta $, $c_{\theta }\equiv \cos \theta $.
{}From (\ref{matnu})-(\ref{angles}) the solar and atmospheric neutrino
oscillation
parameters are
$$
\Delta m^2_{21 }\sim 2m'^2\epsilon\simeq 7\cdot 10^{-11}~{\rm eV}^2~,
$$  
\beq
{\cal A}(\nu_e \to \nu_{\mu , \tau }) =1-{\cal O}(\epsilon^2)
\simeq 0.9 - 1~,
\label{solosc}
\eeq
$$
\Delta m^2_{32}\simeq m_{\nu_3}^2\sim 4\cdot 10^{-3}~{\rm eV}^2~,
$$
\beq
{\cal A}(\nu_{\mu }\to \nu_{\tau })=\frac{4b^2c^2}{(b^2+c^2)^2}-
{\cal O}(\epsilon^2)~,
\label{atmosc}
\eeq
where the oscillation amplitudes are defined as
\beq
{\cal A}(\nu_{\alpha }\to \nu_{\beta })=
4\Sigma_{j<k}U_{\nu }^{\alpha j}U_{\nu }^{\alpha k}
U_{\nu }^{\beta j}U_{\nu }^{\beta k}~.
\label{defamp}
\eeq

We see that the solar neutrino puzzle is explained by maximal vacuum
oscillations of $\nu_e $ into $\nu_{\mu, \tau }$. For $b\sim c$ the
$\nu_{\mu }-\nu_{\tau }$ mixing is naturally large, as suggested by the
atmospheric anomaly. For $b\simeq c$ the $\nu_{\mu }-\nu_{\tau }$ mixing
will be even maximal and $\nu_e $ state oscillations will be $50\%$ into
$\nu_{\mu }$ and $50\%$ into $\nu_{\tau }$. This bi-maximal neutrino
mixing scenario, which we have realized in the framework of SUSY $SO(10)$
model, closely resembles the bi-maximal neutrino mixing suggested
in ref. \cite{bimaxsu5} for $SU(5)$ GUT.  

One may wonder whether the small mixing angle MSW solution 
for the solar neutrino 
puzzle can be realized within our $SO(10)$ scheme. From
(\ref{downlep})
we see that the expected mixing between $\nu_e$ and $\nu_{\mu, \tau }$
states is $\sim \epsilon $, which is too large for the small angle MSW
oscillations. This can be improved if we modify the ${\cal U}(1)$
charge of $10_1$ state to $Q_{10_1}=1/2$. The oscillation angle will
then have the desirable value $\sim \epsilon^2 $. With this modification the
hierarchies between the down quark and charged lepton masses will still
be reasonable, $m_{e, d}/m_{\mu, s}\sim \epsilon^3$. To obtain
$\nu_e-\nu_{\mu, \tau }$ oscillation we can introduce a $SO(10)$ singlet
state $N$ (instead of ${\cal N}_{1, 2}$ states), which will provide mass 
in the $10^{-3}$~eV range to the
`$\nu_2$' state, so that the small angle MSW oscillation for explaining the
solar neutrino 
deficit is realized. 

As far as the large mixing angle MSW solution is concerned, by keeping the 
${\cal N}_{1, 2}$ states with the transformation properties in (\ref{Nu1z2}),
maximal $\nu_e-\nu_{\mu, \tau }$ oscillations will still hold and the 
desired scale ($\sim 10^{-6}~{\rm eV}^2$) can be generated by taking 
$\kappa'\sim 10^{-2}$ in (\ref{scales}).
The oscillation picture (\ref{atmosc})
for the atmospheric neutrinos will be unchanged.

Before concluding, let us briefly discuss the question of nucleon decay within
the proposed $SO(10)$ scheme. First of all, let us note that the operators
$16_i16_j\bar C\bar C'$ and $16_i16_j\bar C\bar C$,
which could induce the dominant decay modes \cite{dommodes} in $SO(10)$
are
forbidden by the ${\cal U}(1)$ symmetry. This means that
the states $\nu^c_{16_i}$ remain massless. However, this does not affect 
anything since they do not have `Dirac'-type couplings with the 
`light' $\nu_{10_i}$ states (the operators $10_i16_j\bar CH$
are forbidden by ${\cal U}(1)$ symmetry).

The couplings $qqT$ and $ql\bar T$ 
emerge from (\ref{up}) and (\ref{downlep}) respectively
and take the form:
\beq
q\hat{Y}_UqT_H+q\hat{Y}_Dl\bar T_{C'}~,
\label{pred5}
\eeq
where ${Y}_U$, ${Y}_D$ denote the Yukawa matrices of up and down quarks, and
$T_H$ and $\bar T_{C'}$ are color triplets from $H$ and $C'$
respectively. In order to build dimension $5$ operators 
the triplet states must be
integrated out. Taking into account equation (\ref{scal1})
as well as the assumption that color triplets
{}from $H$ have masses of order $M_G$,
the nucleon decay amplitude (for $p\to K\nu $) will be suppressed  
$\sim \frac{\epsilon_G}{M_P\epsilon^5}\simeq \frac{1}{5M_G}$,
leading to
a suppression
by a factor $5-10$, relative to the minimal $SU(5)$ scheme. 
We therefore estimate the proton life time to be 
$\tau_{p\to K\nu }[SO(10)]\sim (10-100)\cdot  
\tau_{p\to K\nu}[{\rm minimal}~SU(5)]$.
Hopefully, SuperKamiokande can observe such decays in the not too distant
future!

\vspace{0.3cm}

This work was supported in part by  DOE under Grant No. DE-FG02-91ER40626
and by NATO, contract number CRG-970149.

\vspace*{-0.5cm}

\end{multicols}
\end{document}